\documentclass[aps,twocolumn,notitlepage,superscriptaddress,longbibliography]{revtex4-1}
\usepackage{amsmath}
\usepackage{amssymb}
\usepackage{todonotes}
\usepackage[unicode=true,colorlinks=true,citecolor=blue,urlcolor=blue]{hyperref}

\usepackage{bm}
\usepackage{epsfig}
\usepackage{lipsum}
\usepackage{float}
\usepackage[normalem]{ulem}

\renewcommand {\phi}{{\varphi}}
\newcommand {\rmi}{{\rm i}}

\newcommand {\e}{{\rm e}}
\newcommand {\eps}{\varepsilon}

\begin{document}
\title{
{Classification of three-photon states in waveguide quantum electrodynamics}}

\author{Janet Zhong}
\email{janet.zhong@anu.edu.au}
\affiliation{Nonlinear Physics Centre, Research School of Physics, Australian National University, Canberra ACT 2601, Australia}

\author{Alexander N. Poddubny}

\affiliation{Nonlinear Physics Centre, Research School of Physics, Australian National University, Canberra ACT 2601, Australia}
\affiliation{Ioffe Institute, St. Petersburg 194021, Russia}

\begin{abstract}
We provide the first classification of three-photon eigenstates in a finite periodic array of two-level atoms coupled to a waveguide. We focus on the strongly subwavelength limit and show the hierarchical structure of the eigenstates in the complex plane. The main characteristic eigenstates are explored using entanglement entropy as a distinguishing feature.
We  show that the rich interplay of effects from order, chaos to localisation found in two-photon systems extends naturally to three-photon systems. There also exist interaction-induced localised states unique to three-photon systems such as bound trimers, corner states and trimer edge states.
\end{abstract}
\date{\today}

\maketitle

\section{Introduction} 
A central goal of quantum optics is achieving strong light-matter interactions~\citep{Roy2017,KimbleRMP2018}. This can be realized using cavity quantum electrodynamic (QED) platforms or atomic ensembles~\cite{Chang2019}. A field with growing theoretical and experimental interest is waveguide QED, which is the study of arrays of atoms coupled to a one-dimensional (1D) waveguide. Such systems exhibit collective radiative decay (Dicke super- and subradiance~\citep{Dicke1954}) as well as exotic phenomena unique to the waveguide platform. This includes the fermionisation of subradiant states~\cite{Albrecht2019,Molmer2019}, bound dimer pairs~\citep{Zhang2020subradiant} interaction-induced localisation~\citep{Zhong2020}, self-induced topological phases~\citep{poshakinskiy2020quantum} and quantum chaos~\cite{poshakinskiy2020quantumb}. The important ingredient behind these interesting phenomena is the long-range coupling between atoms found in waveguide QED. It is this that distinguishes waveguide QED from cavity QED and related tight-binding models. Waveguide QED is also promising for many applications in quantum information processing. It can allow us to efficiently generate~\cite{gonzalez-tudela_efficient_2017,paulisch_generation_2018,zhang_heralded_2019-1}, detect~\cite{Malz2020}, slow~\cite{everett_stationary_nodate} and store quantum light~\cite{Leung2012}. It is also useful as a platform for quantum simulators of complex many-body physics~\cite{Nori2009,Nori2014}. Experimental waveguide QED platforms can utilize either natural or artificial atoms and examples of existing setups include cold atoms~\cite{Corzo2019},  defect centers~\cite{Sipahigil2016}, superconducting qubits ~\cite{Astafiev2010,Wang2019,Haroche2020,Browaeys2020,Blais2020,Clerk2020,Carusotto2020} and emerging structures based on exciton-polaritons~\cite{Ghosh2020}.

The many-body effects in waveguide QED are  mostly unexplored, and rich physics can be expected based on recent theoretical results in the two-photon states ~\cite{poshakinskiy2020quantum}. There has been experimental evidence of three-photon bound states (or photonic trimers) in atomic Rydberg setups~\cite{liang2018observation}. Recent theoretical studies have investigated the correlation signatures of a coherent three-photon scattering process in waveguide QED~\cite{chen2020correlation},the dynamics of many-body bound states in chiral waveguide QED~\cite{mahmoodian2020dynamics} and the dissipative losses in three-body systems of strongly interacting photons~\cite{kalinowski2020resonant}. To our knowledge, there has been no work reporting an overview of eigenstates in the three-photon subspace in waveguide QED. Bound states, where multiple particles are stuck together, are an active area of research in many fields such as ultracold atoms~\cite{winkler2006repulsively}. For clarification, the bound states we investigate are when the photons are bound together but are still free to propagate along the waveguide~\cite{Zhang2020subradiant}. This is in contrast to similar studies on atom-photon bound states, where the photon is bound to a lattice site~\cite{calajo2016atom,shi2018effective}. In this paper, we extend previous exotic effects in waveguide QED such as fermionisation~\cite{Albrecht2019}, localisation~\cite{Zhong2020}, bound pairs~\cite{Zhang2020subradiant} and chaos~\cite{poshakinskiy2020quantumb} to three-photon systems for the first time and also predict novel quantum  states, requiring at least three photons.

The paper is structured as follows. In Sec.~\ref{sec:model} we describe the waveguide QED Hamiltonian in the three-photon subspace and show the hierarchical composition of the complex energy spectrum. In Sec.~\ref{sec:entropy}, we focus on the strongly sub-wavelength regime (where the phase between acquired by light propagating between neighbouring qubits is very small at $\varphi=\omega_{0} d/c \approx 0.02\ll 1$) and characterise the main eigenstates using entanglement entropy. In Sec.~\ref{sec:classification} we classify the different types of photon-mediated localisation possible in the three-photon subspace. In Sec.~\ref{sec:exotic}, we numerically verify exotic states (corner and three-photon bound states) which occur at slightly larger phase $\varphi=\omega_{0} d/c \approx 0.2$ and 1 respectively.

\section{Model - Waveguide QED with three photons}
\label{sec:model}
We consider a finite periodic array of $N$ two-level atoms coupled to a 1D waveguide. Under the Markovian approximation, the effective non-Hermitian Hamiltonian for this system is given by~\cite{Ke2019,asenjo2017atom,chang2014,Molmer2019}:
 \begin{align}\label{eq:HM}
 \mathcal H=\sum\limits_{m,n=1}^N H_{m,n}b_{m}^{\dag}b_{n}+\frac{\chi}{2}\sum\limits_{n=1}^Nb_{n}^{\dag} b_{n}^{\dag}b_{n}^{\vphantom{\dag}}b_n^{\vphantom{\dag}}\:,
 \end{align}
where the atomic lattice sites are labelled by indices $m,n=1\ldots N$ and $H_{mn}\equiv\hbar \omega_{0} \delta_{m n} -\rmi\hbar \Gamma_0\e^{\rmi \varphi |m-n|}$. Here, $b_{m}$ are the  annihilation operators for the bosonic  excitations of the qubits and $\varphi = \omega_0 d/c$ is the phase acquired by light between the two neighboring qubits, where $\omega_0$ is the atomic resonance of the qubit and $d$ is the qubit spacing. The parameter $\Gamma_0$ is the radiative decay rate of an individual qubit. In this system, photons become
strongly coupled to atoms and create polaritons, and the $\chi$ term represents the on-site polariton-polariton interaction. A single atom cannot be excited twice (termed as photon blockade) and this is represented mathematically by taking the limit $\chi \rightarrow \infty$.  The imaginary part of the Hamiltonian $H_{mn}$ reflects radiative losses into the waveguide and there exists long-ranged light-induced coupling between distance atoms  described by the term $-\rmi\hbar \Gamma_0\e^{\rmi \varphi |m-n|}$.

Our goal  is to understand the main characteristics of the different kinds of the triple-excited  states $|\Psi\rangle = \sum\psi_{abc} b_a^\dag b_b^\dag b_c^\dag |0\rangle$. We can obtain the eigenstates and eigenvalues $3\epsilon$ by diagonalizing the Hamiltonian Eq.~\eqref{eq:HM} in the subspace of the Hilbert space with three excitations.  Specifically, we  construct the effective three-photon Hamiltonian:
\begin{equation}
H^{(abc)}= H^{(a)} \otimes I^{(b)} \otimes I^{(c)} +
 I^{(a)} \otimes H^{(b)}  \otimes I^{(c)} +
 I^{(a)} \otimes I^{(b)} \otimes H^{(c)}
\end{equation}
as a sum of individual Hamiltonians for first, second and third photon which have superscript labels $a,b$ and $c$ respectively. This can be written explicitly as
\begin{equation}
\begin{split}
H^{(abc)}_{ia,ja;ib,jb;,ic,jc}=\delta_{ib, jb} \delta_{ic, jc} H_{ia,ja}\\+ 
\delta_{ia, ja} \delta_{ic, jc} H_{ib,jb}\\+ 
\delta_{ia, ja} \delta_{ib, jb} H_{ic,jc}\:,
\end{split}
\end{equation}
where $ia,ja,ib,jb,ic,jc=1 \ldots N$. The interaction term is
\begin{equation}
\begin{split}
U^{(abc)}_{ia,ja;ib,jb;,ic,jc}= \bigl(\delta_{ia,ib} \delta_{ja, jb} \delta_{ia,ja} + \delta_{ia,ic} \delta_{ja, jc} \delta_{ia,jc}\\ +\delta_{ib,ic} \delta_{jb, jc} \delta_{ib,jc} \bigr) \chi\:.
\end{split}
\end{equation}
The linear eigenvalue problem to obtain the three-particle excitations is then 
\begin{equation}
\left(H^{(abc)}+\mathcal{U}\right) \Psi=3 \varepsilon \Psi.
\end{equation}
It is also convenient to  transform the Hamiltonian to the basis where bosonic symmetry and photon blockade are  imposed explicitly. Namely, instead of the full basis of $N^3$ states we consider a reduced basis of  $N(N-1)(N-2)/6$ states $\tilde \psi$ where
\begin{equation}
[\tilde{\psi}]_{abc}=[\tilde{\psi}]_{acb}=[\tilde{\psi}]_{bac}=[\tilde{\psi}]_{bca}=[\tilde{\psi}]_{cab}=[\tilde{\psi}]_{cba}=\frac{1}{\sqrt{6}} 
\end{equation}
for $a \neq b \neq c$.

\begin{figure*}[t!]
\centering\includegraphics[width=0.85\textwidth]{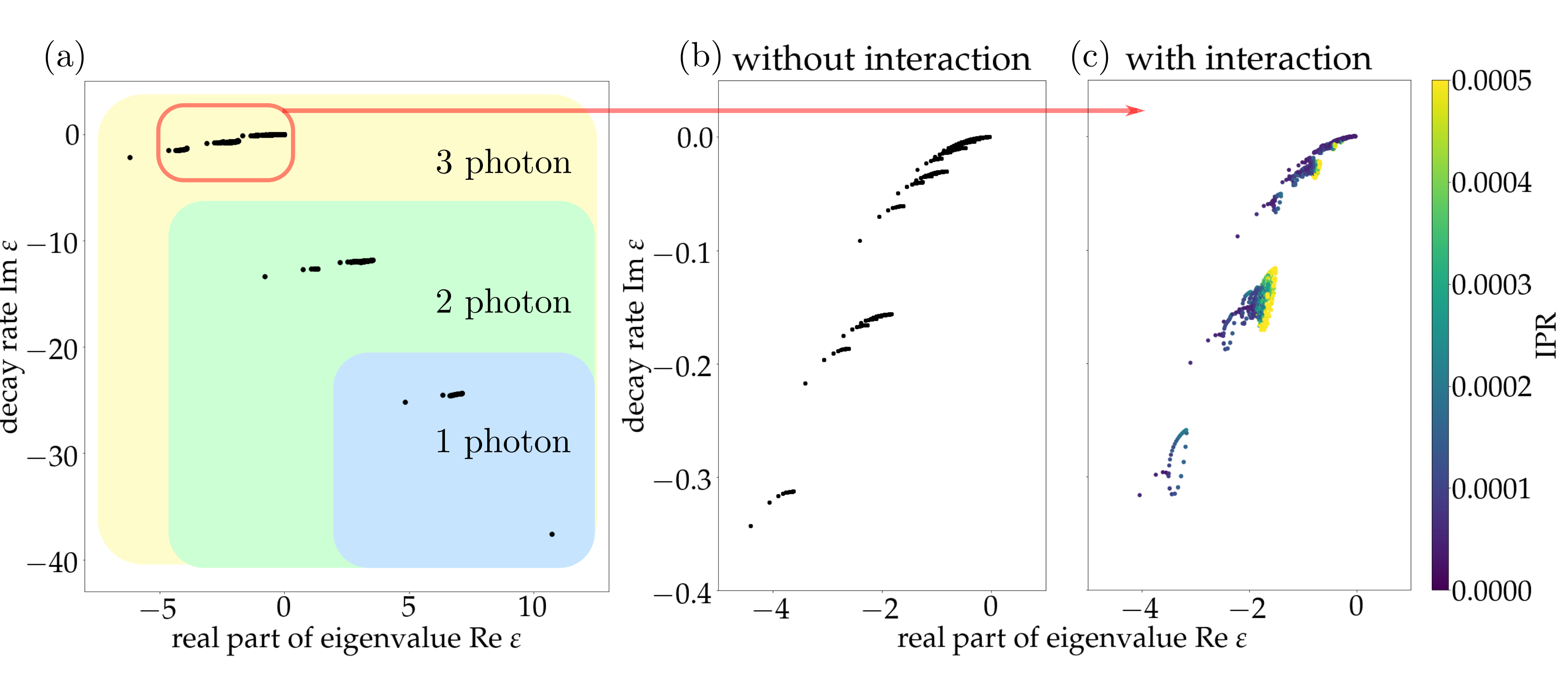}
\caption{(a) Full complex energy spectrum of triple-excited states for an array of $N=42$ atoms coupled to waveguide. (b-c) Eigenvalues correspinding to the red region in (a) at a zoomed scale with $\chi=0$ and $\chi \rightarrow \infty$ respectively. For (c), the eigenvalues are coloured by IPR where higher IPR means it is more strongly localised. Calculation has bee performed for 
$\phi=0.02, \chi \rightarrow \infty$. The energy is measured in the units of $\hbar\Gamma_0$ and counted from the atomic resonance $\hbar\omega_0$.}
\label{figa-complexspectra}
\end{figure*}
In Fig.~\ref{figa-complexspectra}(a), we show the result of numerical calculation of the energy spectrum of a system with $N=42$ atoms and phase $\phi =0.02$ in the three-photon subspace. The eigenvalues are complex because the effective Hamiltonian is non-Hermitian. The imaginary component $\text{Im}(\epsilon) = \Gamma$ of the eigenvalue is the radiative decay rate of the eigenstate. If $\Gamma > \Gamma_0 =-1\mathrm{i}$, where $\Gamma_0$ is the radiative decay rate of a single atom, the eigenstates have an enhanced collective decay rate and are \textit{superradiant}. If $\Gamma < \Gamma_0$, than the collective radiative decay rate is suppressed and the eigenstates are \textit{subradiant}. The structure of the \textit{non-interacting} three photon spectra can be reproduced as just the average of three single photon eigenvalues: $\varepsilon \approx\left(\varepsilon_{a}+\varepsilon_{b}+\varepsilon_{c}\right) / 3$. This means that the structure of the single and two-photon spectra can be seen within the three-photon spectra as shown in the green and blue shaded regions in Fig.~\ref{figa-complexspectra}(a).
In order to understand the hierarchical structure of the spectrum, we take into account that the single photon spectrum $\eps_a$ becomes denser in the region closer to the atomic resonance $\hbar\omega_0$ due to the low group velocity and high density of states [see also the following discussion of Fig.~\ref{fig:fermion}]. Hence, if   $\varepsilon_b$ and $\varepsilon_c$ are fixed to some values far from the resonance, the variation of $\varepsilon_a$ results in a part of the three-photon spectrum that repeats the single-photon spectrum.
 The calculation demonstrates that the interactions leads to only slight modifications of the energy spectrum as a whole. Fig.~\ref{figa-complexspectra}(a) is plotted with infinite on-site repulsion but on this scale, the spectrum looks near indiscernible to a non-interacting spectra. Interaction effects become more visible when zooming in, as seen in the comparison between Fig.~\ref{figa-complexspectra}(b) and Fig.~\ref{figa-complexspectra}(c) which are plotted without and with interaction respectively. In Fig.~\ref{figa-complexspectra}(c), the eigenvalues are coloured by the inverse participation ratio (IPR) of their corresponding eigenvectors. IPR is defined as 
\begin{equation}
\text{IPR}= \frac{\sum_{abc}|\psi_{abc}|^4}{(\sum_{abc}|\psi_{abc}|^2)^2},
\end{equation}
and a high (low) IPR indicates the state is highly localised (delocalised). The high IPR eigenvalues coloured yellow in Fig.~\ref{figa-complexspectra}(c) corresponds to interaction-induced localised states which are discussed in more detail in Sec.~\ref{sec:classification}. The cause for the smearing of eigenstates due to interaction is not fully understood but is in line with numerical results from the two-photon case in Ref.~\cite{poshakinskiy2020quantum}.
\section{Classification of eigenstates via entanglement entropy}
\label{sec:entropy}
\begin{figure}[b]
\includegraphics[width=0.49\textwidth]{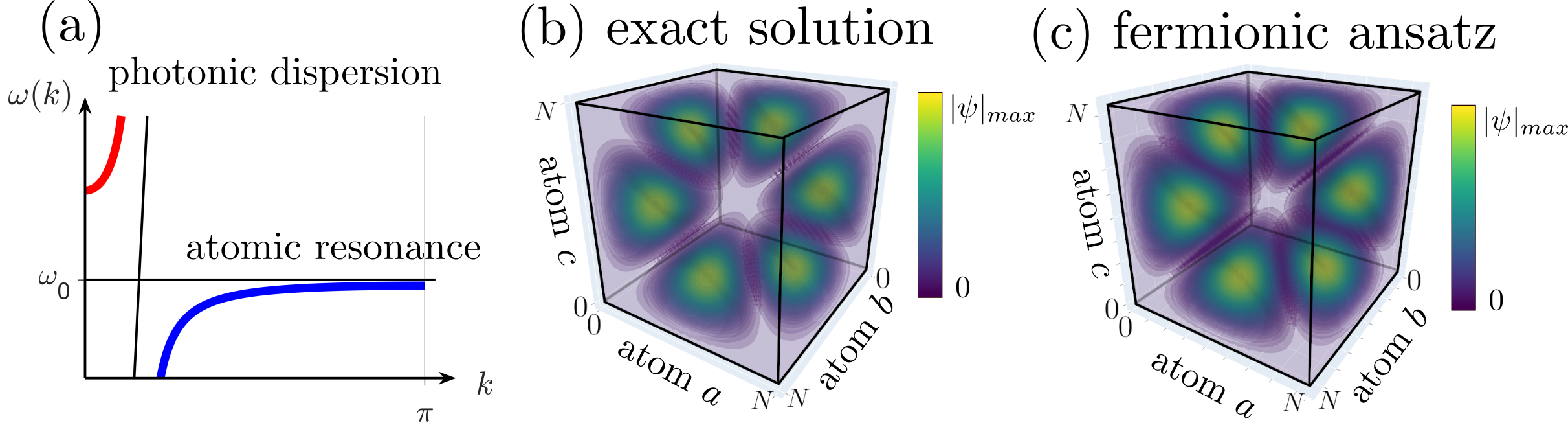}
\caption{(a) Polaritonic dispersion relation for waveguide QED system (b) Fermionic state from exact solution of a system with $N=42$ atoms, three photons and $\omega_0-0.02$. (c) Fermionic ansatz constructed from anti-symmetric combination of three subradiant single-photon eigenstates.}
\label{fig:fermion}
\end{figure}

\begin{figure*}[t!]
\centering\includegraphics[width=0.95\textwidth]{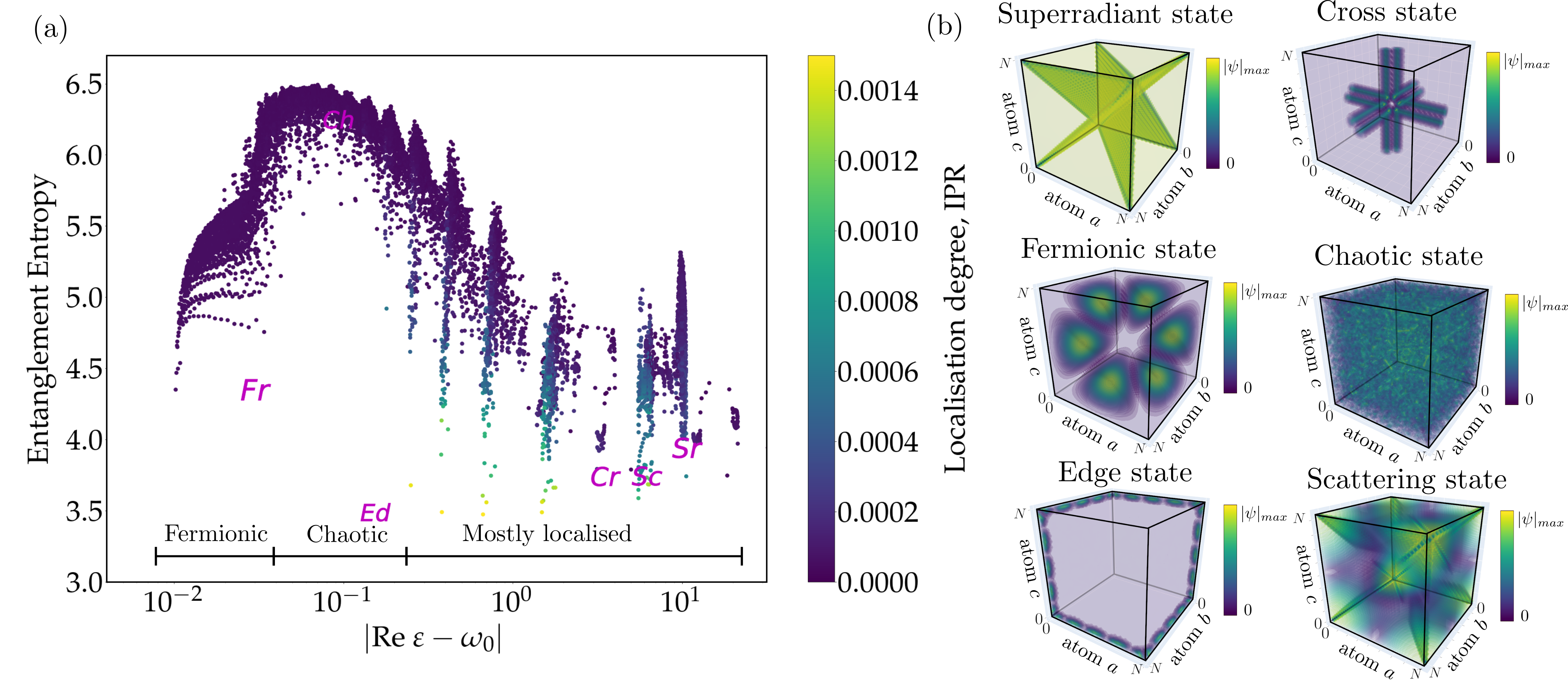}
\caption{(a) Entanglement entropy of three-photon states (b) Characteristic three-photon wavefunctions of superradiant, cross, chaos, edge, fermion  and scattering states which have eigenvalues of $\varepsilon=10.7298-37.58969\mathrm{i}, 3.5414-11.8320\mathrm{i}, -0.0727-0.0004\mathrm{i}, -0.1339-0.0029\mathrm{i},-0.010-2.272 \times 10^{-8}\mathrm{i}$ and $ -6.182-2.156\mathrm{i}$ respectively. Calculation parameters are the same as in Fig.~\ref{figa-complexspectra}.}
\label{fig1-entropy}
\end{figure*}

The tripartite wavefunction can be rewritten using the Schmidt decomposition:
\begin{equation}
\psi_{abc}=\sum_{\nu=1}^{N} \lambda_{\nu} \psi_{a}^{\nu} \psi_{b}^{\nu} \ \psi_{c}^{\nu}\:.
\end{equation}
This is useful because the Schmidt coefficients $ \lambda_{\nu}$ can be used to calculate the von Neumann entropy:
\begin{equation}
S=\frac{-\sum\left|\lambda_{\nu}\right|^{2} \ln \left|\lambda_{\nu}\right|^{2}}{ \sum\left|\lambda_{\nu}\right|^{2}} \:.
\end{equation}
It has previously been shown that the von Neumann entanglement entropy is a useful distinguishing feature for the two-photon subspace~\cite{poshakinskiy2020quantum}, so we will apply this tool to characterise the main eigenstates of the three-photon case. The entanglement entropy of all eigenstates for a system of $N=42$ atoms, three photons and $\omega_0=0.02$ is plotted against the rescaled energy $| \text{Re } \varepsilon - \omega_0|$ in Fig.~\ref{fig1-entropy}(a). There are three main `regions,' of eigenstates, a fermionic, chaotic and a mostly localised region. A few characteristic eigenstates are depicted in Fig.~\ref{fig1-entropy}(b). In these 3D volume plots, the three axes each show the spatial probability of one photon/ polariton along the 1D atomic lattice site index. The probability amplitude is plotted with an opacity of 10\%. All eigenstates in the single-photon subspace are standing waves~\cite{tsoi2008quantum}. The three-photon \textit{scattering state} is simply the symmetrized product of three independent standing waves and is very delocalised. These eigenstates can be better understood from examining the single-particle dispersion relation~\cite{Albrecht2019} $\varepsilon(k)=\Gamma_{0} \sin \varphi /(\cos k-\cos \varphi)$, where $k$ is the Bloch wave vector, so that $\psi_j\propto \e^{\rmi kj}$. This polaritonic dispersion has an upper  ($k<\phi$) and lower ($k>\phi$) branch consisting of an avoided crossing between the photonic dispersion and atomic resonance as depicted in Fig.~\ref{fig:fermion}(a). The scattering states are comprised of three photons in the steeper section of the dispersion closer to the photonic dispersion, where $k$ is on the order of $\varphi$. A steeper curve means the polariton has a smaller effective mass and larger group velocity. Thus, on-site polaritonic repulsions have less time to take effect which results in scattering states resembling non-interacting eigenstates. It is well-approximated by the ansatz:
\begin{equation}
\Psi= c_{123}+c_{321}+c_{213}+c_{312}+c_{132}+c_{231} \:,
\label{eq:symmetricansatz}
\end{equation}
where we define $c_{123} \equiv u_{1}(a) u_{2}(b) u_{3}(c), c_{213} \equiv u_{2}(a) u_{1}(b) u_{3}(c) $ etc. as short-hand notation where $u_{1}, u_{2}, u_{3}$ are constituent single-polariton wave functions and $a,b, c$ are the lattice site indices for the three polaritons. For the scattering states they are standing waves/ single-polariton eigenstates with different wave-vectors. 

The \textit{superradiant state} is also a scattering state, but it corresponds to the upper polaritonic branch. The one depicted in Fig.~\ref{fig1-entropy}(b) has the largest radiative decay rate. The \textit{fermionic state} is  the anti-symmetric combination of subradiant states and can be described by the ansatz
\begin{equation}
\Psi_{a<b<c}= c_{123}-c_{321}+c_{213}+c_{312}-c_{132}-c_{231}  \:.
\label{eq:fermion}
\end{equation} 
When any of the
two particles swap sign, note that Eq.~\eqref{eq:fermion} must be multiplied by $-1$ to preserve bosonic symmetry (the magnitude of the wave-function remains unchanged).
Here, $u_{1}, u_{2}, u_{3}$ are again single-polariton eigenstates. The fermionic state can be understood as three polaritons in the flatter region of the polaritonic dispersion closer to $k=0$ or $\pi$. Flatter dispersion means a slower group velocity. This lends more time for interactions to play a role and the repulsive on-site interaction results in a `Pauli-exclusion' effect~\cite{Albrecht2019}. The anti-symmetric ansatz is plotted in Fig.~\ref{fig:fermion}(c) and it well matches the exact eigenstate depicted in Fig.~\ref{fig:fermion}(b).

The \textit{cross states} and the \textit{edge states} are interaction-induced localisation states. They are described by an ansatz of the form of Eq.~\eqref{eq:symmetricansatz} where $u_{1}, u_{2}, u_{3}$ are either $u_{\text{edge}}, u_{\text{centre}}$ or $u_{\text{free}}$ depending on the number of photons localised (see Sec.~\ref{sec:classification} for more details). $u_{\text{edge}},u_{\text{centre}}$ are single-polariton wave-functions localised at the edge or centre of the atomic array respectively but they are \textit{not} single-particle eigenstates. $u_{\text{free}}$ is a free, delocalised standing wave that closely resembles a single-polariton eigenstate. This localisation is a novel phenomena unique to waveguide QED and is caused by light polaritons being trapped in the nodes of the standing waves of heavy polaritons. It can be shown that long-range interactions are crucial for this localisation effect~\cite{Zhong2020}. The fact that the localisation is interaction-induced distinguishes it from more common localisation effects such as Anderson localisation.

 The \textit{chaotic state} cannot be decomposed into a few single particle states~\cite{poshakinskiy2020quantumb}. It is characterised by a highly irregular wave-function in real space and high entanglement entropy which supports the notion that it is comprised of many entangled single-particle states.  Recently, a study on analogous states in two-photon systems~\cite{poshakinskiy2020quantumb} shows that they cannot be described using the Bethe ansatz. This suggests that the problem is non-integrable and exhibits quantum chaos. These chaotic states cannot be described by an equation in the form of Eq.~\eqref{eq:symmetricansatz}.

\section{Classification of localisation}
\label{sec:classification}
\begin{figure*}[t!]
\centering\includegraphics[width=0.62\textwidth]{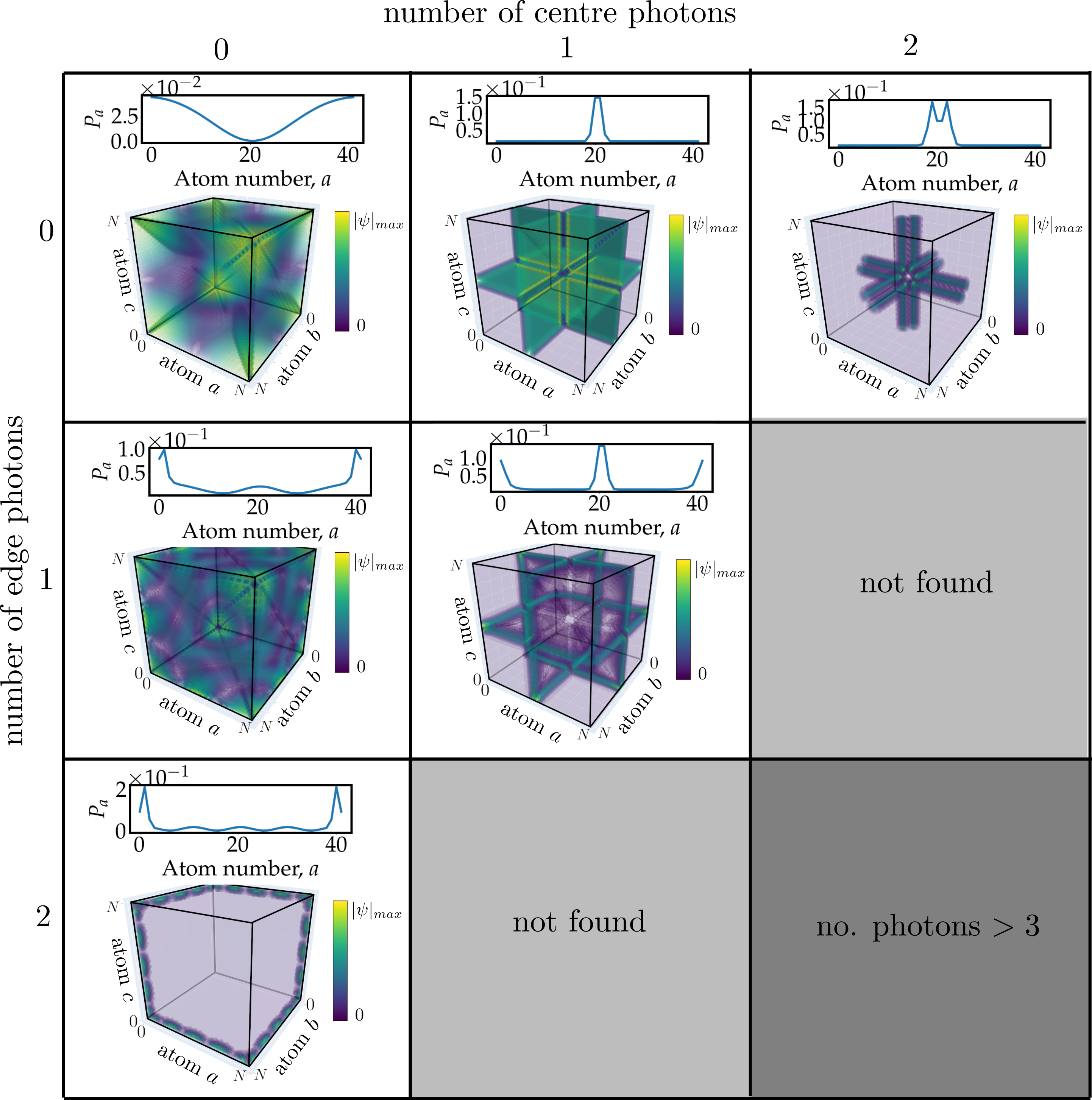}
\caption{Classification of the different types of localisation possible according to number of photons localised at the edge or centre of the qubit array. Above each 3D probabilty density plot, we show a 1D representation of the probability amplitude defined by $P_a$ as given in Eq.~\eqref{eq:pa}. If the sum of photons from the table does not add to three, then the other photons are free, delocalised photons.}
\label{fig-localisation}
\end{figure*}

We study the interaction-induced localisation in waveguide QED~\cite{Zhong2020} in more detail for a three-photon subspace. A motivating question is how many photons we can have localized in the three photon subspace. In Fig.~\ref{fig-localisation}, we classify the types of localisation possible according to whether the photons are localised at the centre of the qubit array or at the edge. We numerically verify that we can have combinations of 0,1 or 2 photons localised at either the edge or the centre of the atom array. Note that we can also have localisation that is away from the centre and the edge. Our ansatz is of the form of Eq.~\eqref{eq:symmetricansatz}:
\begin{equation}
\Psi \approx u_{1} u_{2} u_{3} + \text{(bosonic symmetry terms)},
\label{ansatz}
\end{equation}
where $u_{1}, u_{2}, u_{3}$ are either $u_{\text{edge}}, u_{\text{centre}}$ or $u_{\text{free}}$ depending on number of photons localised as according to Fig.~\ref{fig-localisation} (if the photon is not localised it is free) where the bosonic symmetry terms are just the permutations of the first term in Eq.~\ref{ansatz}. We have manually searched to verify that the maximum number of photons we can have localised in this small subwavelength limit is two photons, as classified in Fig.~\ref{fig-localisation}. Having one photon localised corresponds to a plane of high probability amplitude, having two photons localised corresponds to a line and having three photons localised would correspond to a point in the 3D probability density plot. It may be possible to achieve a situation where all three photons are localised at higher phase, which is briefly mentioned in Sec.~\ref{sec:exotic}, but eigenstates at higher phase are vastly unexplored even in the two-photon case. In Fig.~\ref{fig-localisation}, we also plot a 1D representation of the probability amplitude which we have coined $P_a$. This is defined as 
\begin{equation}
P_a = \sum_{b,c} \psi_{abc}^2
\label{eq:pa}
\end{equation}
where $\psi_{abc}$ is a component of the three-photon probability amplitude. This just allows us to see where the polariton is localised, as it may be difficult to discern what is happening inside the 3D probability density plots from just the volume plots. We sum the square because otherwise it can sometimes cancel out if parts of the wave-function is localised but of opposite parity. We see in this 1D representation that the localisation at either the edge or centre is quite apparent.\\

\section{Exotic states at larger distance between atoms}
\label{sec:exotic}
\begin{figure}[t]
\centering
\includegraphics[width=0.35\textwidth]{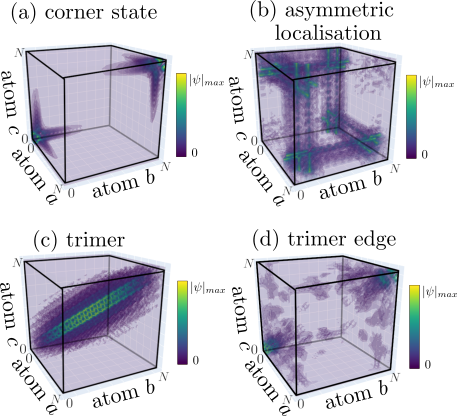}
\caption{Exotic three-photon states for larger distance between atoms $d$. The corresponding phases $\phi=\omega_0d/c$ are equal to $\phi=1, \pi+0.3$ and $0.2$ for the the trimer and asymmetric localisation, trimer edge and corner state respectively. Other parameters are the same as in Fig.~\ref{figa-complexspectra}.}
\label{fig:exotic}
\end{figure}
In this Section, we go beyond the strongly subwavelength limit to look for exotic states at higher distance between the atoms, characterized by the phase $\varphi=\omega_0d/c$. In particular, we aim to numerically verify the existence of trimers (three-photon bound state) in waveguide QED systems. Subradiant dimers in the two-photon subspace have been predicted theoretically quite recently, only in 2019~\cite{Zhang2020subradiant}. Photonic trimers were experimentally observed in 2017 in a Rydberg polariton setup but to the best of our knowledge has not been studied in a waveguide QED model of periodic array of two-level atoms coupled to a waveguide. For the phase $\phi=1$, we numerically verify an existence of  several trimers states. The trimer corresponds to the three photons being bound to each other, which means that the probability amplitude exponentially decays from the main diagonal of the 3D probability density plot as shown in Fig.~\ref{fig:exotic}(c).

Unlike the two-photon case, we can also have asymmetrical (relative to the positions in the 1D qubit array) localisation for three photons as depicted in Fig.~\ref{fig:exotic}(b). With two photons, the standing wave potential must be symmetrical and our localisation can only be symmetrical for a periodic lattice. When the lattice geometry is no longer periodic/ symmetrical, we can have asymmetrical localisation, as studied in the modulated spaced qubit array in Ref.~\cite{Poshakinskiy2014}. In the three-photon case, since we can have overlapping polaritons, this can break the symmetry of the self-induced potential which allows for asymmetrical localisation. The asymmetrical localisation also occurs at small phase (such as $\phi\approx 0.02$ but can become even more prevalent at higher phase. Interestingly, at just a slightly higher phase of $\phi=0.2$ we can get corner states, as shown in Fig.~\ref{fig:exotic}(a). This is when the photon is exponentially decaying from the edge. This could be related to topological edge states or even higher order topological edge states but this is beyond the scope of this paper.

Finally, we also searched for trimer edge states which is when the three photons are bound together and also localised at the edge. This can be seen as a three-photon extension of the radiative bound pair edge states studied in Ref.~\cite{Ke2020}. We find some states that satisfy this description at a phase $\phi=\pi+0.3$ and plot this in Fig.~\ref{fig:exotic}(d). The difference between these states and corner states is that they are exponentially decaying from the main diagonal of the 3D volume plot whereas corner states are exponentially decay along the edges of the volume plot. 
It is unclear at the moment whether the localization of these  trimer edge states has any  topological origin. Many exotic effects at higher phase are not understood even at the two-photon level, however the very fact that these states exist shows the incredible richness of these waveguide QED systems. We have not yet found dimer states in the three photon subspace, where only two photons of three are bound to each other. 

\section{Summary and Outlook}
Within this work, we give an overview to the broad plethora of states available in the three photon subspace of waveguide QED systems. We showed the hierarchical structure of the complex eigenstates and broadly classified the main types of eigenstates using entanglement entropy. The fermionic, superradiant, chaotic and localised states found in two-photon systems~\cite{Molmer2019,Zhong2020,poshakinskiy2020quantum} extend quite naturally to the three photon case. We present a general classification of the  possible types of interaction-induced localisation possible in the three-photon system. We also numerically verify the existence of exotic states at higher distance between the atoms. Notably, we find bound photonic trimers which generalizes the subradiant dimers predicted  in the two-photon case~\cite{Zhang2020subradiant}. Having three photons also breaks the symmetry of the self-induced polaritonic potential, and so we can get asymmetrical localisation. Contrary to the systems where the asymmetry is embedded in the lattice geometry~\cite{Poshakinskiy2014},
here the asymmetry is self-induced by the interactions.  Detailed structure of bound trimers requires further studies. For example, it might be useful to calculate their dispersion relations depending on the center-of-mass wave vector  as has been done for the bi-photon states  in Refs.~\cite{Zhang2020subradiant,Ke2020}. 
Our numerical results also indicate  a signature of trimer edge state, that generalizes  bound pair edge states discussed in~\cite{Ke2020}. Whether these exotic trimer states are relevant to topological edge states or even higher order topological edge states~\cite{Schindler2018} may be an avenue of future exploration. \\

\let\oldaddcontentsline\addcontentsline
\renewcommand{\addcontentsline}[3]{}

\begin{acknowledgments}
This work was supported by the Australian Research Council under the grant FT170100331. J.Z was supported by the Australian Government Research Training Program (RTP) Scholarship.
\end{acknowledgments}



%

\end{document}